# Voltage synchronizations between multichannel electroencephalograms during epileptic seizures

Çağlar Tuncay
caglart@metu.edu.tr

**Abstract**: The underlying dynamics for the electroencephalographic (EEG) recordings from humans but especially epilepsy patients are usually not completely known. However, the ictal activity is claimed to be characterized by synchronous oscillations in the brain voltages in the literature. These time dependent interdependencies (synchronization, coupling) between the EEG voltages from epileptogenic and non epileptogenic brain sites of nineteen focal epileptic patients are investigated in this work. It is found that strong synchronization-desynchronization events occur in alternation during most of the investigated seizures. Thus, these seizures are detected with considerable sensitivity (71 of the 79 seizures).

**1. Introduction**: Automatic detection of epileptic seizures is a challenging subject in quantified electroencephalography where the underlying dynamics for the recordings, but especially from epileptic patients, are usually not completely known [1 and 2]. Fortunately, the ictal state is claimed to be characterized by synchronous oscillations in the brain voltages which were investigated by various methods in several papers. For example, in [3 and 4] the intracranial EEG data from human subjects with epilepsy were analyzed. It is reported in both publications that there is strong evidence for non-linear interdependence between the focus of epileptic activity and other brain regions and these synchronizations can be detected for up to 30 seconds prior to a seizure. Different researchers [5] studied single electrode scalp EEG recordings from 60 healthy humans and found that non-linear structure is only present weakly and infrequently. Recently, brain synchronizations were tested by comparing 30 measures in terms of their ability to distinguish between the inter-ictal period and the pre-seizure period [6 and 7]. One may refer to [7-10] for the mathematical methods used in the above mentioned approaches based on amplitude, time or frequency domain measurements.

In this work, amplitude analyses based on moving time window technique [11 and 12] will be performed on intracerebrally recorded electroencephalograms from nineteen humans with focal epilepsies. A measure will be defined for the duration of strong synchronizations between channels in each time window and the time trajectories of those metrics will be investigated with the basic aim of detecting seizure onsets.

**1.i Methods**: The EEG voltages ($X_k(i)$) downloaded from the data pool [13] have been recorded from six brain sites (k=1-6) of each patient with integer values expressed in micro Volt (µV), which range from -10,000 µV to 10,000 µV or broader. Obviously, if the sampling rate of ($X_k(i)$) is f Hertz (Hz), then the real time (t) for the univariate is t=i/f; more precisely, ($X_k(i) \rightarrow X_k(t)$) where t=i/f. The investigated data in this study have been recorded with a sampling rate of 256 Hz (see Section 1.ii).

**1.i-a Voltage translation**: The electroencephalographic data, especially from the epilepsy patients, are known to display various spiky behaviors with large amplitudes. Moreover, these spikes may occur frequently in several epochs and then the amplitude synchronizations may be altered unduly. Therefore, the spiky behaviors should be filtered out, which would mean that data are manipulated. An alternative method is used for reducing the effect of spikes on the results in this work: Before the synchronization investigations, the raw data are segmented into time windows with length of ($\Delta\tau=$) 5/256 seconds. More precisely, only 5 data points are included per window at this stage.

These segments can be enumerated by the integer M=1,2,3 … and the sample number (N) per window equals N=($\Delta\tau$)f=5, where

$$f(M-1)\Delta\tau < i \leq fM(\Delta\tau) \quad \text{or} \quad (M-1)\Delta\tau < t \leq M(\Delta\tau) \quad . \tag{1}$$

As a result, a univariate segment of EEG data for a given k (=1-6) in the time window (M) may be represented as $X_{k;M}(i')$ or $X_{k;M}(t')$, where t'=i'/f and

$$0 < i' \leq N \quad \text{or} \quad 0 < t' \leq \Delta\tau \tag{2}$$

for each M. This segment has the following sample mean ($\Lambda_{k;M}$)

$$\Lambda_{k;M} = \sum_{i'=1; N} X_{k;M}(i') / N \tag{3}$$

where N=5 and thus $\Delta\tau$= 5/256 seconds at this stage.

In the second step of the process, the sample mean (Eq. (3)) of each window is subtracted from the EEG voltages $X_{k;M}(i')$. Thus, the EEG voltages ($X_{k;M}(i')$) in each window (M) are translated by the sample mean ($\Lambda_{k;M}$) of that window:

$$x_{k;M}(i') = X_{k;M}(i') - \Lambda_{k;M} \tag{4}$$

with i'≤5 as in Eq. (2) so that the new sample means of the data in each window are close to zero. In this process, spikes are not filtered out, but their disturbing effects on the synchronizations are reduced.

After translating the EEG voltages, each channel will be segmented into consecutive sections of a time length of $\Delta\tau$=2 seconds. In other words, the synchronization analyses, which will be discussed in the following section, will be performed in a time window starting at around ($\tau$) and ending at around ($\tau+\Delta\tau$) with $\Delta\tau$=2 seconds. Next, this time window will be moved stepwise to produce continuously updated spectra of the durations of the strong synchronizations.

**1.i-b Voltage synchronization measure**: Spatiotemporal synchronizations between two segments in the window M can be detected as follows: Consider the translated voltages from two channels $x_{k;M}(t')$ and $x_{k';M}(t')$. If the simultaneous spatial differences of the voltages ($x_{k;M}(t') - x_{k';M}(t')$) always occur around zero in the window, then $x_{k;M}(t')$ and $x_{k';M}(t')$ are said to be synchronized throughout the corresponding epoch. Furthermore, the time durations of the synchronizations between bivariate segments can be estimated in terms of the number of the times (event frequency) for

$$x_{k;M}(t') - x_{k';M}(t') \sim 0 \quad . \tag{5}$$

These durations will be divided by the time length of the window ($\Delta\tau$) in order to obtain a parameter between 0 and 1, as a measure of durations of the voltage synchronizations in the given time window. This time window will be moved stepwise to produce continuously updated spectra of synchronization measures. These measures are henceforth referred to as the voltage synchronization measure (VSM) indices and designated by the symbol $V_{k,k';M}$ or $V_{k,k'}(\tau)$. Obviously, the longest duration corresponding to $V_{k,k'}(\tau)=1$ is equal to $\Delta\tau$ seconds. As a result, the investigations to understand the underlying mechanisms of the synchronizations entail the analysis of the simultaneous spatial differences of brain voltages in this work. Note that the models used are time invariant as stipulated in [11].

**1.ii Material**: Recordings from twenty one patients are available in the data pool [13] where it is stated that, "The EEG database contains invasive EEG recordings of the patients suffering from medically intractable focal epilepsy." Moreover, "to obtain a high signal-to-noise ratio, fewer artifacts, and to record directly from focal areas, intracranial grid-, strip, and depth-electrodes were utilized. The EEG data were acquired using a Neurofile NT digital video EEG system with 128 channels, 256 Hz sampling rate, and a 16 bit analogue-to-digital converter. Notch or band pass filters have not been applied.

For each of the patients, there are data sets called 'ictal' and 'inter-ictal', the former containing files with epileptic seizures and at least 50 minute pre-ictal data. The latter contains approximately 24 hour (h) of EEG-recordings without seizure activity. At least 24 h of continuous interictal recordings are available for 13 patients. For the remaining patients interictal invasive EEG data consisting of less than 24 h were joined together, to end up with at least 24 h per patient. For each patient, the recordings of three focal and three extra-focal electrode contacts is available." The technical details of the recordings are summarized in Table 1.

This pool of data has been downloaded by global research groups and treated with several aims [7, 14 and 15].

The data recorded from epileptogenic zones will be designated with (k=1-3) and those from non epileptogenic ones with (k=4-6), here. As a result, data recorded from each of the six contact positions (k) in ictal interval or inter-ictal interval of a patient are segmented into consecutive sections of length of $\Delta\tau=2$ seconds, which amounts to 512 data per window since the sampling rate (f) is 256 Hz. Then, the indices will be calculated for the data in each of the non overlapping windows.

The results are presented in the following section. The last section is devoted to discussion and conclusion.

**2. Results**: Six time trajectories of $V_{k,k'}(\tau)$ will be investigated in this section; more precisely, $V_{1,2}(\tau)$, $V_{2,3}(\tau)$ and $V_{3,1}(\tau)$ for the in-focus electrodes and $V_{4,5}(\tau)$, $V_{5,6}(\tau)$ and $V_{6,4}(\tau)$ for the out-focus electrodes. The seizure time terms of each patient are noted in terms of integers in [13] and these numbers are utilized to select the seizure time windows. Moreover, the data from two patients, Pat01 and Pat02 as referred in [13], will be disregarded since the seizure time terms are not long enough to study with two-second-long windows and the inter-ictal data are missing, respectively. Furthermore, a threshold of 10 μV is assumed to define the strong synchronizations between the bivariate segments in this work. To be more precise, strong synchronizations can be said to have been settled between the voltages $x_{k;M}(t')$ and $x_{k';M}(t')$ if $|x_{k;M}(t') - x_{k';M}(t')| \leq 10$ μV. The empirical distributions of the VSM metrics obtained from all of the available data from nineteen patients under that assumption are presented in Figure1, where bin size used is 0.001.

On inspection of Fig.1, it can be observed that the data from focal epileptic brain sites are strongly synchronized in nearly one tenth of numerous two-second intervals since the maximum values of the event frequency histograms (counts) occur commonly at around 0.1. It can be assumed that, this result is valid for the time intervals longer than two seconds. The same result may further be assumed for the healthy brains since most of the data used are recorded from the inter-ictal intervals and half of those are from the non epileptogenic brain areas. However, the mentioned rates may be altered during epileptic seizures or other brain complications as will be discussed in further paragraphs. It is worth noting that the behavior for the large or small metrics are different for the in-focus and out-focus electrodes. Moreover, the noisy behavior for large metrics from the out-focus contacts is found to be patient-specific.

Another relevant indication may be that the lines for the counts from the in-focus electrodes follow each other more closely with respect to those from the out-focus electrodes. This result may be expected since the in-focus electrode sites are closely located at the nucleus of the epileptogenic zones and the out-focus electrodes are located far from the nucleus; i.e., at the non epileptogenic region, which is wider. In brief, the VSM histograms are found to depict characteristics as well as patient-specific features.

**2.i VSM time profiles around seizures**: The behavior of the in-focus and out-focus VSM time trajectories around an investigated seizure is depicted in Figure 2 where the time-axis for ($\tau$) does not represent the real time but designates the time domain of the events. On inspection of Fig.2, it can be observed that the VSMs attain small values during the seizure which is designated with solid squares at the bottom part of the figure. This behavior indicates short durations for strong synchronizations and thus can be treated as desynchronization as well. It may be claimed that desynchronization between the channels is the relevant feature for the seizures of several (if not all of the) patients, and hence, this feature can be used for algorithmic or statistical seizure detection; see, Section 2.ii.

The collective phenomenon between the channels may be another important observation on Fig.2 since the VSM time trajectories from the in-focus and out-focus electrodes descend collectively during the seizure. To be more specific, the time profiles of $V_{1,2}$ and $V_{5,6}$ or $V_{2,3}$ and $V_{3,1}$, or $V_{4,5}$ and $V_{6,4}$ are close to 0.1 before the onset. However, they collectively attain small values during the seizure. Furthermore, the profiles (especially $V_{1,2}$ and $V_{5,6}$) again collectively ascend towards 1 after the offset. Later, a transition time term occurs for the profiles to descend towards the intermediate values; i.e., those around 0.1.

However, not all of the above discussed aspects of the profiles can be observed during the entirety of the investigated seizure time terms. For example, the out-focus electrodes do not always couple with the in-focus electrodes during seizures, or the profiles may not ascend towards 1 at the times close to seizure offsets. Nevertheless, desynchronization process, which lasts longer than four seconds; i.e., covering at least two successive windows, is the common aspect of the VSM time trajectories during several seizures.

**2.ii Detecting seizures**: As a result, the model proposed for the detection of the seizure time terms of the investigated nineteen patients relies upon one of the following criteria:

  1) At least two simultaneous VSM time trajectories from the in-focus electrodes should attain values smaller than 0.05 within two consecutive time windows.
  2) These features which show decrease first and then depict horizontal behavior covering at least two consecutive time windows and then increase towards the intermediate values or higher are selected in the theoretical data.

It should be noted that, only the recordings from at least two in-focus electrodes are concerned in the applications of the given criteria, where the extreme values should cover at least two successive time windows; thus, the spontaneous occurrences can be disregarded. Moreover, various SVM time profiles are found to attain higher values than 0.1 prior to different seizures. In the occasion of this case, the extreme values may not fall below the threshold used for the first criterion and thus, that seizure can not be detected. Furthermore, the profiles from the in-focus contacts should descend collectively below the pre-seizure levels within the assumed characteristic feature for the second criterion, where no threshold is concerned.

When the first criterion is applied to the data, 65 of the 79 seizures are detected with the success rate of 82.3%. At the same time, the total rate of false positive results is 3.92 per hour. However, it is important to note that the majority (89.9%) of those false positive values come from only 5 patients (Pat04, Pat 06, Pat09, Pat14, and Pat19).

When only the second criterion is applied, the results are better; 71 seizures are detected with the success rate 89.9%. Furthermore, the false positive rate also drops considerably to 0.06 per hour.

**3. Discussion and conclusion**:   It is clear that the closeness of the means of the time windows with five data points to zero does not warrant the closeness of the means of the two-second-long windows to zero. In addition, the spiky behaviors, whose problematic features have been mentioned in Section 1.i-a, were not filtered out in the data used. These reasons may constitute the background for obtaining less than 100 % sensitivity in this work. One may refer to Table 1 in [7] for comparison of different methods where the best rates reported are 100 % success rate and zero false positive per hour.

It is worth underlining that the proposed model is new. Moreover, the model possesses the potential to become routine practice since it is based on channel synchrony. In other words, it is statistically shown that strong synchronization and desynchronization events in alternation are important for the occurrences of nearly 90% of the investigated seizures. Thus, it is claimed that the electrographic features for strong synchronization and desynchronization events are due to electrophysiological coupling and decoupling of the epileptogenic areas.

On inspection of the theoretical data obtained in this work, it is observed that the ictal intervals and inter-ictal intervals can be distinguished for some patients by means of various oscillations in the SVM time profiles as exemplified in Fig.2. In addition, the initiation of the pre-seizure states as stressed in [6 and 7] can be distinguished in a similar way. Thus, the inter-ictal intervals and pre-seizure states can be anticipated for the prediction of the seizure onsets, several minutes before the occurrences. However, these discussions would be made in a lengthy and more comprehensive report.

**ACKNOWLEDGEMENT**
The author is thankful to the University of Freiburg for their kindness in giving permission to investigate their database and use all of the related material. A special thank is devoted to Mr. Murat Aydin for his friendly helps and discussions.

**FIGURES**

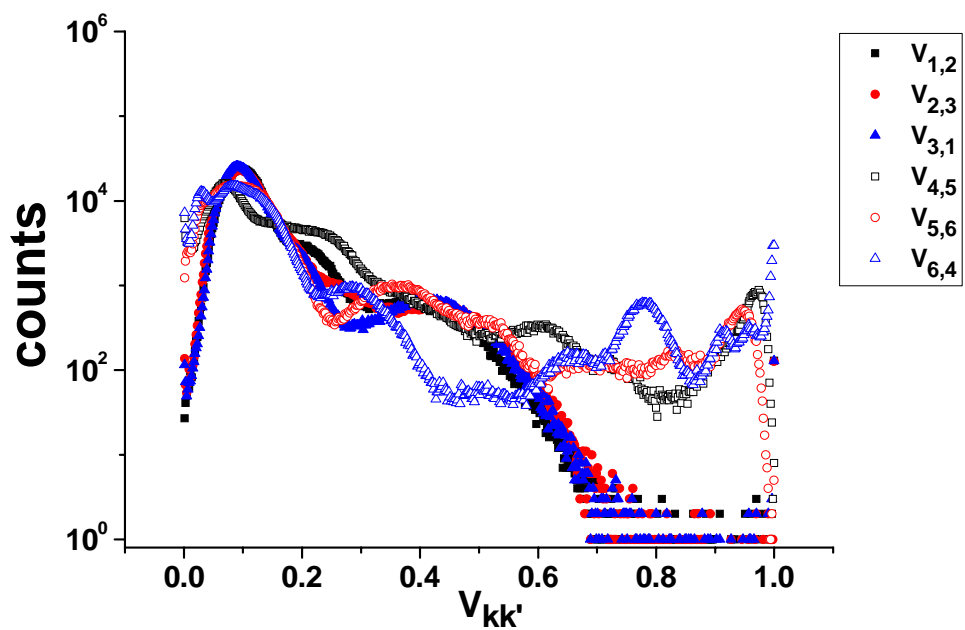

**Figure 1** The empirical distributions of the VSM metrics obtained from all of the available data.

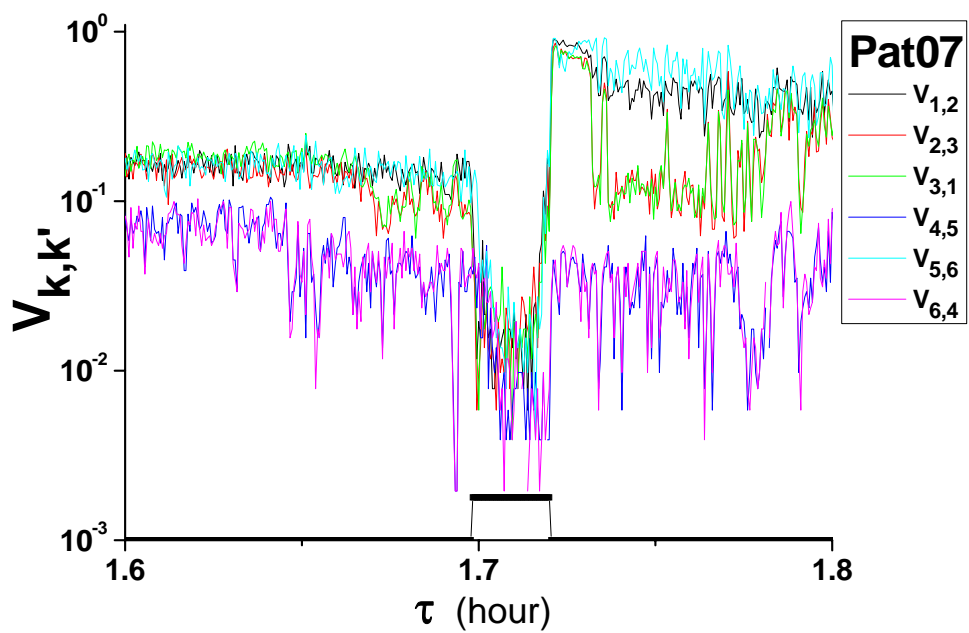

**Figure 2** Oscillations in the time courses of the VSMs from Pat07 around a seizure.

**Table 1**     In this table, patients' genders, ages, seizure types, their origin of epilepsy, electrodes types applied, and the number of seizures are given. Eight of the patients are male, and thirteen are female. The seizure types are simple partial (SP), complex partial (CP), and generalized tonic-clonic (GTC). Epileptic foci are neocortical (NC) for eleven patients, hippocampus (H) for eight of them, and both for two patients. The recordings were taken by intracranial grid (g), strip (s) and depth (d) electrodes, not all of which were used for each patient.

| Patient | Sex | Age | Seizure type | Epileptic focus | Origin of epilepsy | Electrodes | Seizures analyzed |
|---|---|---|---|---|---|---|---|
| 1 | F | 15 | SP,CP | NC | Frontal | g,s | 4 |
| 2 | M | 38 | SP,CP,GTC | H | Temporal | d | 3 |
| 3 | M | 14 | SP,CP | NC | Frontal | g,s | 5 |
| 4 | F | 26 | SP,CP,GTC | H | Temporal | d,g,s | 5 |
| 5 | F | 16 | SP,CP,GTC | NC | Frontal | g,s | 5 |
| 6 | F | 31 | CP,GTC | H | Temporo/Occipital | d,g,s | 3 |
| 7 | F | 42 | SP,CP,GTC | H | Temporal | d | 3 |
| 8 | F | 32 | SP,CP | NC | Frontal | g,s | 2 |
| 9 | M | 44 | CP,GTC | NC | Temporo/Occipital | g,s | 5 |
| 10 | M | 47 | SP,CP,GTC | H | Temporal | d | 5 |
| 11 | F | 10 | SP,CP,GTC | NC | Parietal | g,s | 4 |
| 12 | F | 42 | SP,CP,GTC | H | Temporal | d,g,s | 4 |
| 13 | F | 22 | SP,CP,GTC | H | Temporo/Occipital | d,s | 2 |
| 14 | F | 41 | CP,GTC | H, NC | Fronto/Temporal | d,s | 4 |
| 15 | M | 31 | SP,CP,GTC | H, NC | Temporal | d,s | 4 |
| 16 | F | 50 | SP,CP,GTC | H | Temporal | d,s | 5 |
| 17 | M | 28 | SP,CP,GTC | NC | Temporal | s | 5 |
| 18 | F | 25 | SP,CP | NC | Frontal | s | 5 |
| 19 | F | 28 | SP,CP,GTC | NC | Frontal | s | 4 |
| 20 | M | 33 | SP,CP,GTC | NC | Tempo/Parietal | d,g,s | 5 |
| 21 | M | 13 | SP,CP | NC | Temporal | g,s | 5 |